\documentclass[twoside,a4paper,11pt,twocolumn]{article}
\usepackage{extsizes}
\usepackage[super,sort&compress,comma]{natbib} 
\usepackage[left=1.5cm, right=1.5cm, top=1.785cm, bottom=2.0cm]{geometry}
\usepackage{times,mathptmx}
\usepackage{sectsty}
\usepackage{graphicx} 
\usepackage{lastpage}
\usepackage[format=plain,justification=raggedright,singlelinecheck=false,font={stretch=1.125,footnotesize,sf},labelfont=bf,labelsep=space]{caption}
\usepackage{fancyhdr}
\usepackage{fnpos}
\usepackage[english]{babel}
\usepackage{array}
\usepackage{gensymb}
\usepackage{droidsans}
\usepackage{charter}
\usepackage[utf8]{inputenc}
\usepackage[T1]{fontenc}
\usepackage[usenames,dvipsnames]{xcolor}
\usepackage{setspace}
\usepackage[compact]{titlesec}
\usepackage{textgreek}
\usepackage{url}
\usepackage{dblfloatfix} 
\usepackage{tabu}
\usepackage{amsmath, eqnarray}
\usepackage{dsfont}
\usepackage{upgreek}
\usepackage{soul}
\usepackage[switch]{lineno}
\usepackage[symbol]{footmisc}

\usepackage[hidelinks]{hyperref}

\newcommand{\beginsupplement}{%
        \setcounter{table}{0}
        \renewcommand{\thetable}{S\arabic{table}}%
        \setcounter{figure}{0}
        \renewcommand{\thefigure}{S\arabic{figure}}%
     }

\title{Boundary shape engineering for the\\ spatial control of confined active particles}

\author{
Roberto Di Leonardo\textit{$^{1,2,7}$}
András Búzás,\textit{$^{3}$} 
Lóránd Kelemen,\textit{$^{3}$}\\
Dávid Tóth,\textit{$^{4,5}$}
Szilvia Z. Tóth,\textit{$^{4}$}
Pál Ormos,\textit{$^{3}$} 
Gaszton Vizsnyiczai,\textit{$^{3,6,7}$}
}

\date{\small 
\textit{$^{1}$~Dipartimento di Fisica, Sapienza Università di Roma, Piazzale A. Moro 5, I-00185, Rome, Italy}\\
\textit{$^{2}$~NANOTEC-CNR, Soft and Living Matter Laboratory, Institute of Nanotechnology, Piazzale A. Moro 5, 00185, Rome, Italy}\\
\textit{$^{3}$~Institute of Biophysics, HUN‑REN Biological Research Centre Szeged, Temesvári krt. 62, H-6726, Szeged, Hungary}\\
\textit{$^{4}$~Laboratory for Molecular Photobioenergetics, Institute of Plant Biology, HUN‑REN Biological Research Centre, Szeged, Temesvári krt 62, H-6726 Szeged, Hungary}\\
\textit{$^{5}$~Doctoral School of Biology, University of Szeged, Közép fasor 52, H-6722 Szeged, Hungary}\\
\textit{$^{6}$~Department of Biotechnology, University of Szeged, Közép fasor 52, H-6726, Szeged, Hungary}\\
\textit{$^{7}$These authors jointly supervised this work.}\\
Correspondence and requests for materials should be addressed to R.D.L and V.G. (email: roberto.dileonardo@uniroma1.it, vizsnyiczai.gaszton@brc.hu)
}

\begin{document}
\pagestyle{fancy}
\thispagestyle{plain}
\twocolumn[
  \begin{@twocolumnfalse}

\maketitle
\begin{abstract}
\noindent 
Unlike an equilibrium gas, the distribution of active particles can be very sensitive to what happens at the boundaries of their container. Experiments and simulations have previously highlighted the possibility of exploiting this behavior for the geometric control of active particles, although a general theoretical framework is lacking.
Here we propose a boundary method based on the flux transfer formalism typical of radiometry problems, where surface elements transmit and receive "rays" of active particles with infinite persistence length. As in the case of blackbody radiation, a Lambert scattering law results in a uniform distribution of active particles within the cavity, while other scattering laws result in specific patterns of particle accumulation in the bulk or over the boundary walls.
We validate our method's predictions with numerical simulations and demonstrate its practical utility by spatially controlling swimming microalgae confined in light-defined arenas.
The presented boundary method offers a simple and efficient way to predict particle distributions when both the geometry of the boundaries and the scattering law are known. In addition, it provides a general design principle for engineering container shapes optimized for transport, accumulation, and sorting of self-propelled colloids and microorganisms. 


\end{abstract}

\end{@twocolumnfalse} \vspace{0.4cm}]
 
\section*{Introduction}

An equilibrium gas will fill any container with a homogeneous density regardless of its shape and material. Conversely, non-equilibrium gases of self-propelled particles are extremely sensitive to boundary geometry and interactions. Asymmetric walls can produce currents and give rise to spontaneous accumulation \cite{galajda2007wall, wan2008rectification, lambert2010collective, katuri2018directed}, while the mechanical pressure exerted by an active gas over container walls can be strongly influenced by the details of interactions \cite{solon2015pressure, pellicciotta2023colloidal}. When active particles collide with a wall, the irreversible nature of self-propulsion results in scattering laws that break time reversal symmetry \cite{di2010bacterial, reichhardt2017ratchet}. Microswimmers like \textit{E. coli} or catalytic Janus particles always end up to be aligned along the wall surface regardless of the incoming angle \cite{bianchi2017holographic, simmchen2016topographical}. While ciliary contact interactions cause puller swimmers such as the single-celled algae \textit{C. reinhardtii} to leave the surface at a constant angle, losing memory of incidence \cite{kantsler2013ciliary, lushi_scattering_2017, thery2021rebound}. This microscopic irreversibility has been exploited to design structured surfaces that could trap, sort, repel and rectify the motion of microswimmers \cite{di2017deployable, hulme2008using,  mok2019geometric, palacios2021guided, galajda2007wall}.
However, thanks to a rich sensory apparatus, the concept of boundary for living systems is broader than just physical walls. Chemotactic bacteria are pushed out of regions where repellents are present, while photosensitive microorganisms can bounce back at abrupt transitions between light and dark \cite{lam2017device}.
In this context, there is a need for a new modeling frameworks that could connect the active particle's boundary scattering characteristics and the confinement's geometry to the stationary distribution of active particles. On the fundamental side this represents a rich class of new problems for the statistical mechanics of active particles. On the practical side this could guide the engineering of boundary shapes to control the spatial distribution of active particles in the bulk, transport and segregate cell types based on wall interactions, or simply inhibit unwanted accumulation for biofilm prevention.


Here we focus on a class of active particles that move along straight trajectories and undergo memoryless scattering processes at the boundary of an arbitrarily shaped region. Using a formalism borrowed from radiometry, we show that the problem of finding the steady state spatial distribution of these particles can be cast into a boundary problem to be solved numerically. 
To verify the correctness of our model, we study how different scattering laws produce different density modulations depending on the shape of the boundary and compare these theoretical results with numerical simulations of active particles. In many equilibrium situations, like photons in a blackbody cavity or rarefied molecules in a Knudsen gas \cite{steckelmacher1986knudsen}, scattering follows a Lambertian cosine law, which in the case of active particles also results in perfectly uniform distributions inside cavities of any shape. Deviations from Lambert's law result in regions of high concentration that shift from the bulk to the boundary as the scattering angles change from normal to tangential directions. In addition we provide a direct experimental application of our method to a system of flagellated microalgae \textit{Euglena gracilis} swimming in optically defined arenas.  Upon encountering a light-dark interface, light-responsive Euglena cells undergo random scattering events that reorient the cells toward the cavity interior and keep them confined. We found that Euglena scatters from light-dark interfaces with a nearly Lambertian law so that cells spread almost uniformly inside simple cavity shapes. Nevertheless, using the results of our boundary method, we were able to design a stacked multi-stage structure that results in a three-fold concentration of Euglena cells between its two ends. Model predictions are quantitatively confirmed by experiments, demonstrating the accuracy of our method and its practical applications.

\section*{Results}

{\bf A boundary element method for confined active particles}.

\begin{figure}[h]
    \includegraphics[width=.5\textwidth]{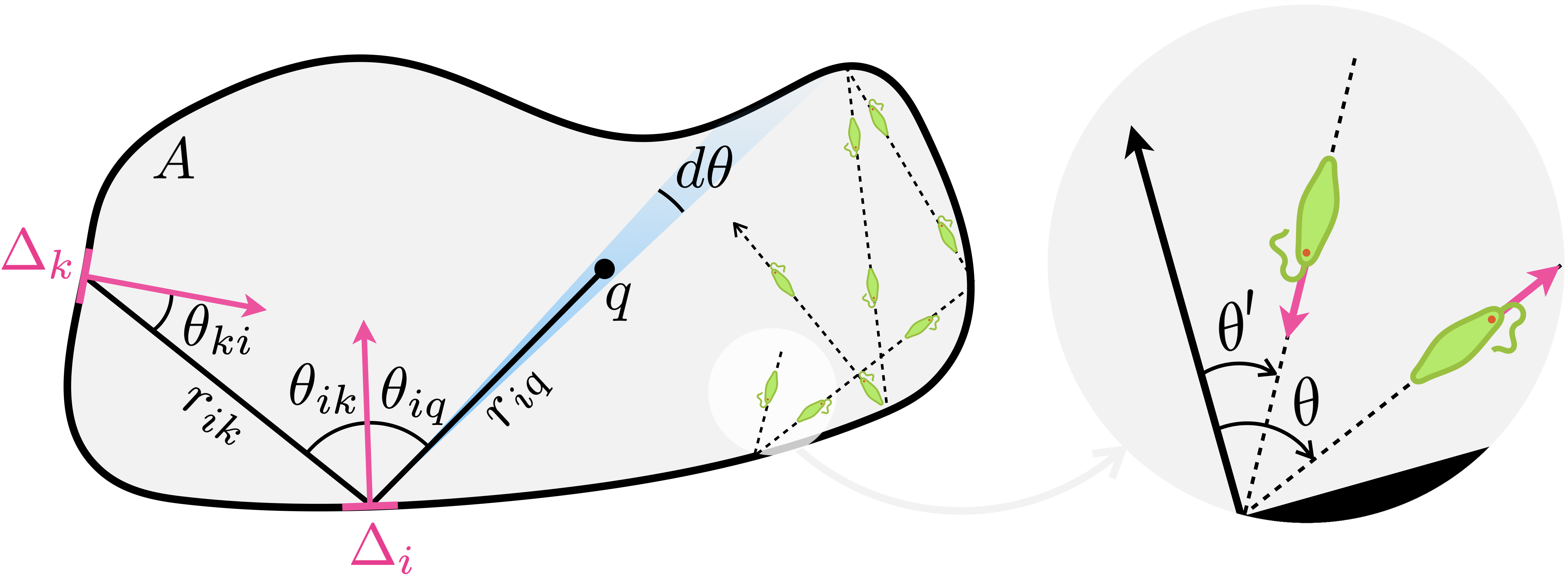}
    \caption{Geometric description of the boundary model.}
    \label{fig:f1}
\end{figure}

When the persistence length of active particles exceeds the size of the container, they will travel in straight lines from one element of the boundary to the next. Every time a particle ``feels'' a boundary through mechanical, hydrodynamical, chemical or optical signals it quickly reorients to a new swimming direction. We call $S(\theta^\prime, \theta)$ the scattering law representing the probability density of being scattered to a new direction $\theta$ given the incident angle $\theta^\prime$. We will assume that the boundary is impenetrable so that both $\theta$ and $\theta^\prime$, defined as in Fig.\ref{fig:f1}, 
vary between $-90^\circ$ and $90^\circ$. 

\begin{figure*}[hb]
    \centering
 \includegraphics[width=.65\textwidth]{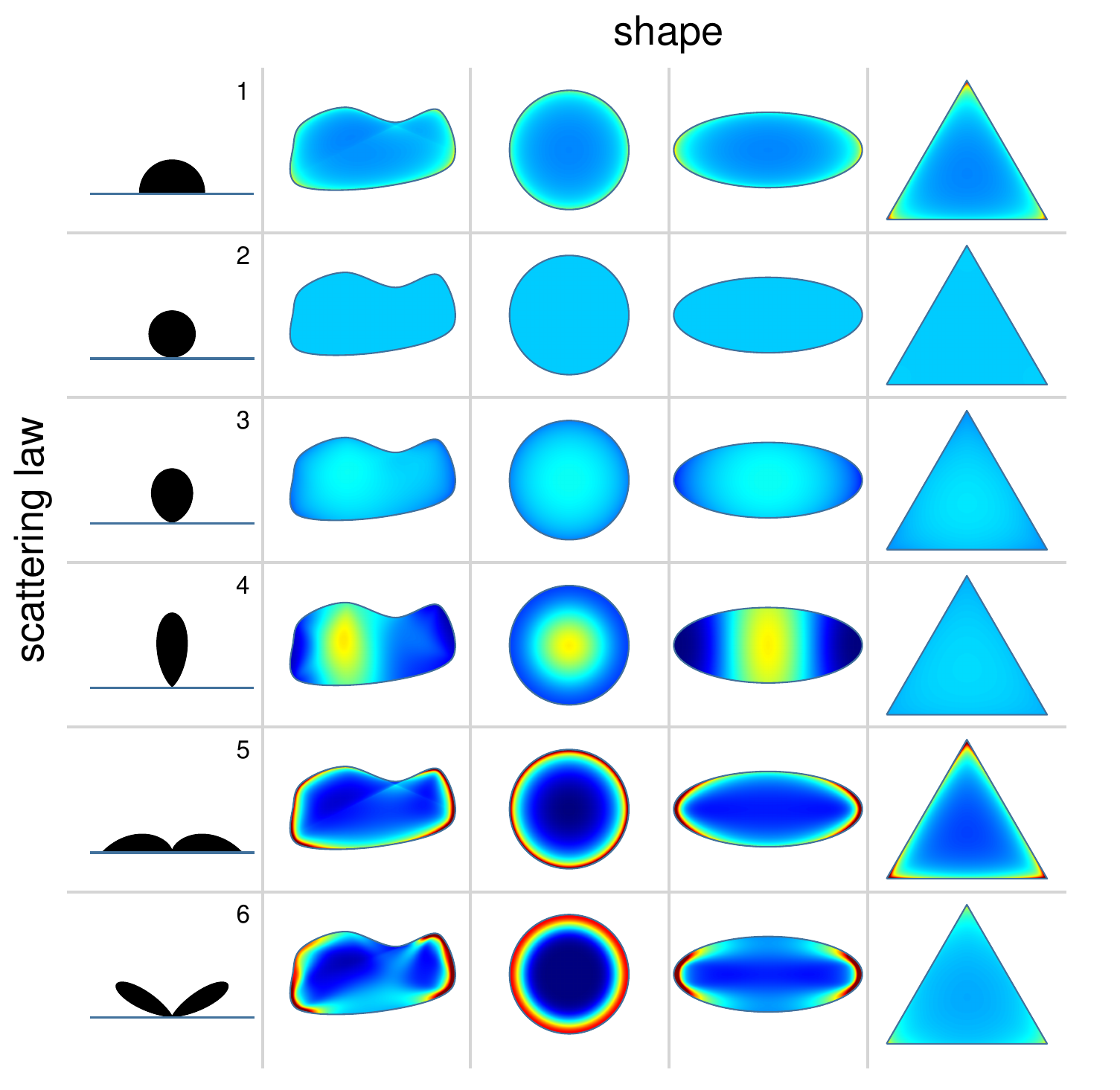}
 \caption{Spatial probability density distributions calculated with the boundary method in four different confinement geometries for the set of scattering angle distributions defined in Table \ref{fig:ftable}.}
 \label{fig:f2} 
\end{figure*}

Then we discretize the boundary in finite elements of lengths $\Delta_i$ and call $J_i(\theta) d\theta$ the stationary flux of particles emerging from the $i$-th boundary element within an angle $d\theta$ centered around $\theta$. This flux results from scattered particles arriving in $\Delta_i$ from all other elements $\Delta_k$ so that we can write:
\begin{equation}
\label{eq:master}
J_i(\theta) \Delta_i = \sum_{k\neq i} J_k(\theta_{ki})\Delta_k \frac{\Delta_i \cos\theta_{ik}}{r_{ik}} S(\theta_{ik},\theta)
\end{equation}
Although, in principle, the above equation could be solved for $J_i(\theta)$ for a given scattering law $S(\theta^\prime, \theta)$ (for instance by discretizing $J_i(\theta)$ into an array of values for equally spaced angles between $-90^\circ$ and $90^\circ$ and so turning it into a linear algebra problem), the solution simplify noticeably when active particles loose memory during the scattering event so that $S(\theta^\prime, \theta)=\sigma(\theta)$. Substituting in Eq.\ref{eq:master} we get:
\begin{equation}
    J_i(\theta) \Delta_i = \sigma(\theta)\sum_{k\neq i} J_k(\theta_{ki})\Delta_k\Delta_i \frac{\cos\theta_{ik}}{r_{ik}}=\sigma(\theta)\Delta_i p_i
\end{equation}
with
\begin{equation}
    \label{eq:linalg}
    p_i \Delta_i= \sum_{k\neq i} J_k(\theta_{ki})\Delta_k \Delta_i \frac{\cos\theta_{ik}}{r_{ik}}=
    \sum_{k\neq i} W_{ik} p_k \Delta_k
\end{equation}
where we introduced the variables $p_i$ representing the stationary flux of particles colliding/emerging on boundary element $i$ per unit time and unit length. The matrix $W_{ik}=\sigma(\theta_{ki})\Delta_i \cos\theta_{ik}/r_{ik}$ represents the probability that a particle scattered from boundary element $k$ will hit next at $i$ so that it must satisfy the normalization rules $\sum_i W_{ik}=1$. The matrix $W_{ik}$ is fully determined by the geometry of the boundary and the scattering law. 
In concave boundaries the connection of segment pairs can be occluded, for which case $W_{ik}$ must be set to $0$.  
Once these are fixed the $p_i$ values can be found by solving the linear algebra problem in Eq.\ref{eq:linalg} that in matrix form reads:
\begin{equation}
    \label{eq:system}
        \begin{pmatrix}
        -\Delta_{1} & W_{12}\:\Delta_{2} & \cdots & W_{1n}\:\Delta_{n}\\
        W_{21}\:\Delta_{1} & -\Delta_{2} & \cdots  & W_{2n}\:\Delta_{n}\\
        \vdots & \vdots & \ddots & \vdots\\
        W_{n1}\:\Delta_{1} & W_{n2}\:\Delta_{2} & \cdots & -\Delta_{n}
        \end{pmatrix}
        \cdot
        \begin{pmatrix}
        p_{1}\\
        p_{2}\\
        \vdots\\
        p_{n}
        \end{pmatrix}
        =
        \begin{pmatrix}
        0\\
        0\\
        \vdots\\
        0
        \end{pmatrix}
\end{equation}    
Once the boundary fluxes $p_i$ have been determined the particle density at a generic internal point $q$ with coordinates $\mathbf r = (x,y)$ can be obtained as the sum of density contributions from all elements:
\begin{equation}
    \label{eq:rho}
        \rho(\mathbf r)=\sum_{i=1}^{n} p_{i} \Delta_{i} \frac{\sigma(\theta_{iq})}{v r_{iq}}  
\end{equation}
where $v$ is the particle speed. A possible way of seeing it is shown in Fig.\ref{fig:f1} where $p_i\Delta_i \sigma(\theta_{iq})d\theta$ represents the flow of particle emerging from element $\Delta_i$ within an angle $d\theta$ around the direction $\theta_{iq}$. This flow can be also written as $\rho_i(\mathbf r) v\;r_{iq}\;d\theta$ where $\rho_i(\mathbf r)$ is the density of particle emitted from $\Delta_i$.  Form that we find $\rho_i(\mathbf r)=p_{i} \Delta_{i} \sigma(\theta_{iq})/v r_{iq}$ and summing over all elements we get the total density in Eq.\ref{eq:rho}. The values $p_i$ derived from Eq.\ref{eq:system} are defined within an arbitrary multiplicative factor so that we can absorb the factor $v$ in the $p_i$ values and normalize in the end by imposing the condition $\int_A \rho(\mathbf r)d^2 r=1$ where $A$ is the surface enclosed by the boundary. 

\noindent{\bf The role of geometry and scattering law}. 

As a first application of our boundary method we investigate how the combined effects of boundary shape and scattering function affect the spatial distribution of active particles in the interior. We choose four different geometries: a generic shape, a circle, an ellipse and a equilateral triangle. For every geometry we consider the six different scattering laws:
\begin{table}[h]
    \centering
    \includegraphics[width=.35\textwidth]{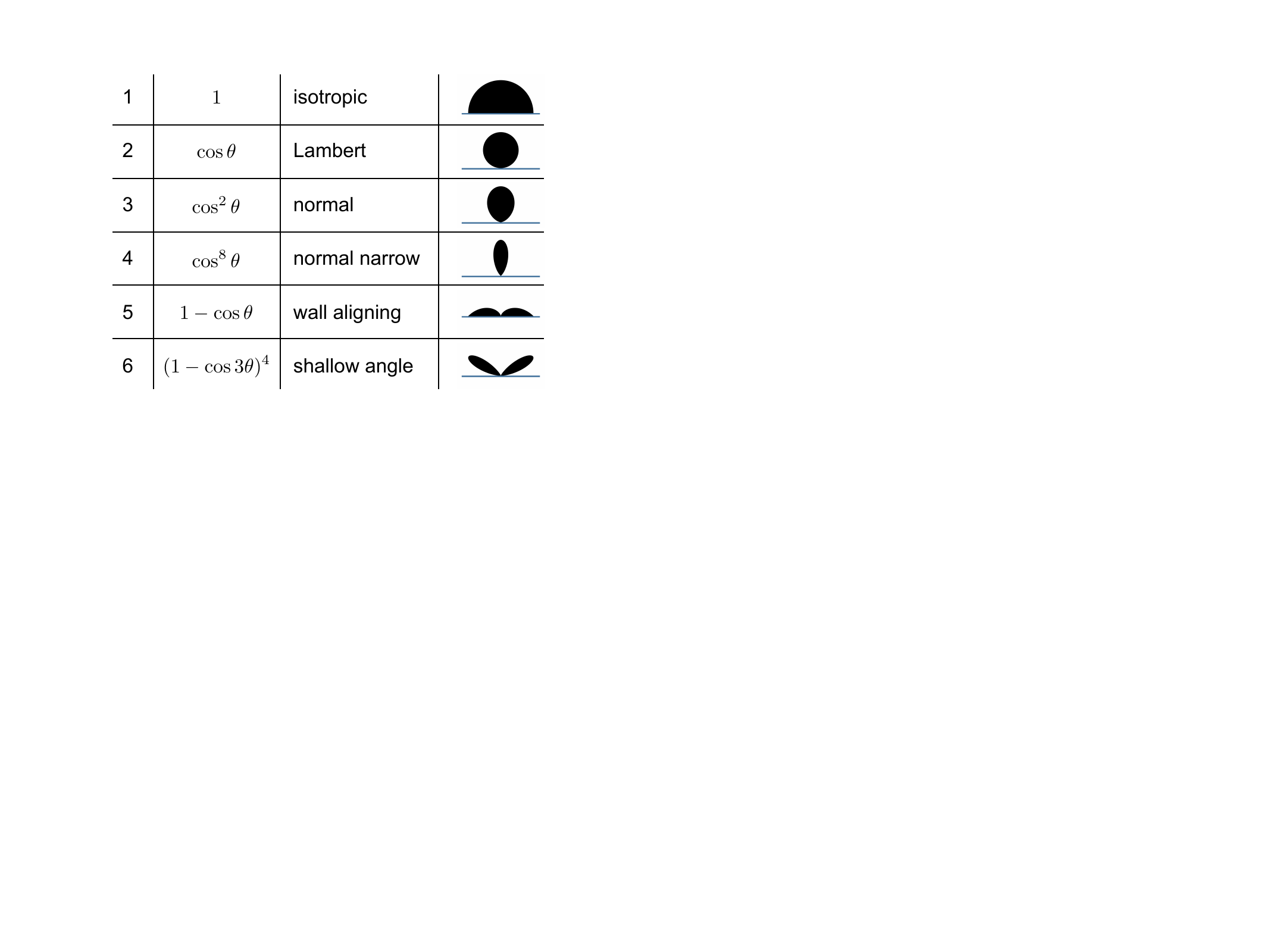}
    \caption{List of the examined scattering laws.}
    \label{fig:ftable}
\end{table}
The first one represents isotropic scattering, particles emerge with an angle that is uniformly distributed between -90 and 90 degrees. The second, third and fourth scattering laws describe cases where particles emerge with an increasingly narrow distribution around the surface normal, represented mathematically as the cosine of the outgoing angle raised to increasingly larger powers. The fifth case describes particles that are preferentially aligned parallel to the wall while the last one corresponds to a shallow scattering angle as found for \textit{Chlamydomonas} algae scattering from solid surfaces.


The spatial distribution of active particles in the cavity is strongly dependent on both geometry and scattering law (Fig.\ref{fig:f2}). 
As expected, the wall-aligning and the shallow angle scattering result in a high probability of finding active particles close to the boundary, favouring regions of higher convex curvature. These results for shallow angle scattering in the circle and ellipse confinements match with the previously reported experimental observations of single \textit{Chlamydomonas} microalga cells in such compartments \cite{ostapenko2018curvature, cammann2021emergent}. 
Wall accumulation is a general feature of active particles and is often the consequence of slowing down due to wall repulsion counteracting self-propulsion \cite{bechinger2016active}. But here we are considering instantaneous scattering events with particles that do not slow down when they hit the boundary, so it is somewhat surprising to find this wall accumulation even for isotropic scattering. 
In contrast, scattering distributions that favour outgoing directions close to the surface normal tend to generate concentration peaks in the inner region, while maintaining a low density at the edge. Interestingly, in the equilateral triangle very narrow and normal directed scattering results in an almost uniform distribution.



Another notable exception is when the scattering probability is in the form of $\cos{\theta}$ (number 2 in the table), which produces a uniform concentration for any geometry. Interestingly, this kind of particle scattering is analogous to Lambert's cosine law for photons, where the cosine shaped scattering makes the brightness of Lambertian  surfaces invariant to the observation angle. It also characterises the emission of radiation from the surface of a blackbody cavity, a consequence of the isotropic radiation (uniform energy density) inside the cavity. In terms of our boundary model, when the scattering law is $\sigma(\theta)=\cos\theta/2$ then we have the detailed balance condition:

$$
W_{ik} \Delta_k=\frac{\cos\theta_{ki}\cos\theta_{ik}\Delta_i\Delta_k}{2r_{ik}}=W_{ki}\Delta_i
$$
so that Eq.\ref{eq:linalg} admits the constant solution $p_i=p^*$

\begin{equation}
    \label{eq:linalg_lambert}
    p^* \Delta_i=
    \sum_{k\neq i} W_{ik} p^* \Delta_k=
    \sum_{k\neq i} W_{ki} p^* \Delta_i=
    p^* \Delta_i\sum_{k\neq i} W_{ki}
\end{equation}
where the identity follows from the normalization condition $\sum_{k\neq i} W_{ki}=1$. A uniform boundary flux $p^*$ produces a uniform bulk density $\rho(\mathbf r)$ as can be easily deduced from Eq.\ref{eq:rho} noting that for Lambert scattering $\Delta_i \cos\theta_{iq}/r_{iq}$ is just the angular size of element $\Delta_i$ from point $q$.

\begin{figure}[h]
    \centering
    \includegraphics[width=.5\textwidth]{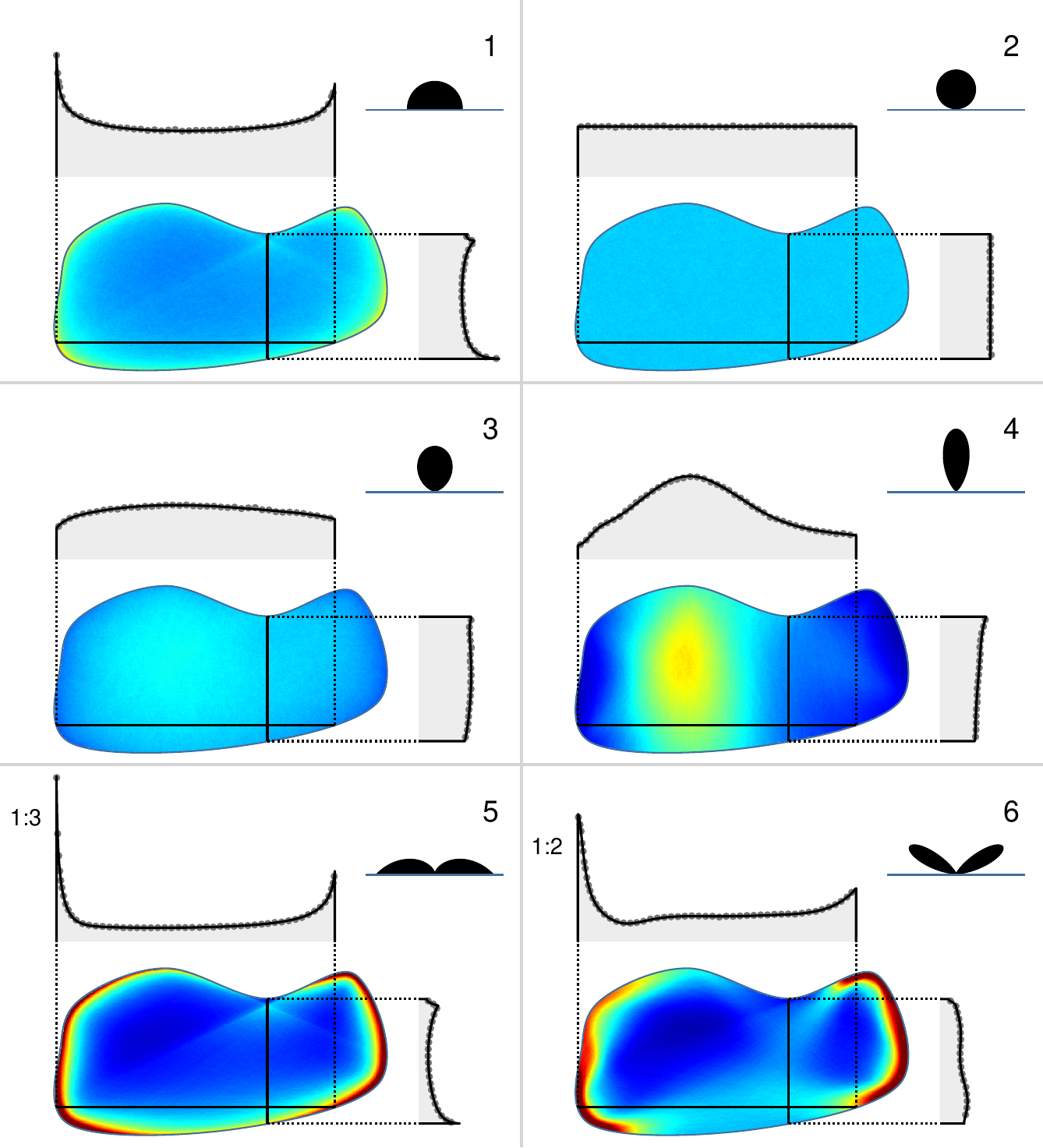}
    \caption{Validation of the boundary method with active particle simulations. Two dimensional spatial probability density distributions obtained with particle simulations are shown for a generic shape and a set of scattering laws. Results of the two methods are compared on plots of density profiles along two sample lines, with the simulation results represented by gray circles and the boundary model results by black lines.}
    \label{fig:f3} 
   \end{figure}

To validate the results of our method we have performed numerical active particle simulations (see Methods) whose results are reported for a selected geometry in Fig.\ref{fig:f3}. As shown by density profiles along two sample lines the agreement between theory and simulation is perfect.

\begin{figure*}[t]
    \centering
    \includegraphics[width=.85\textwidth]{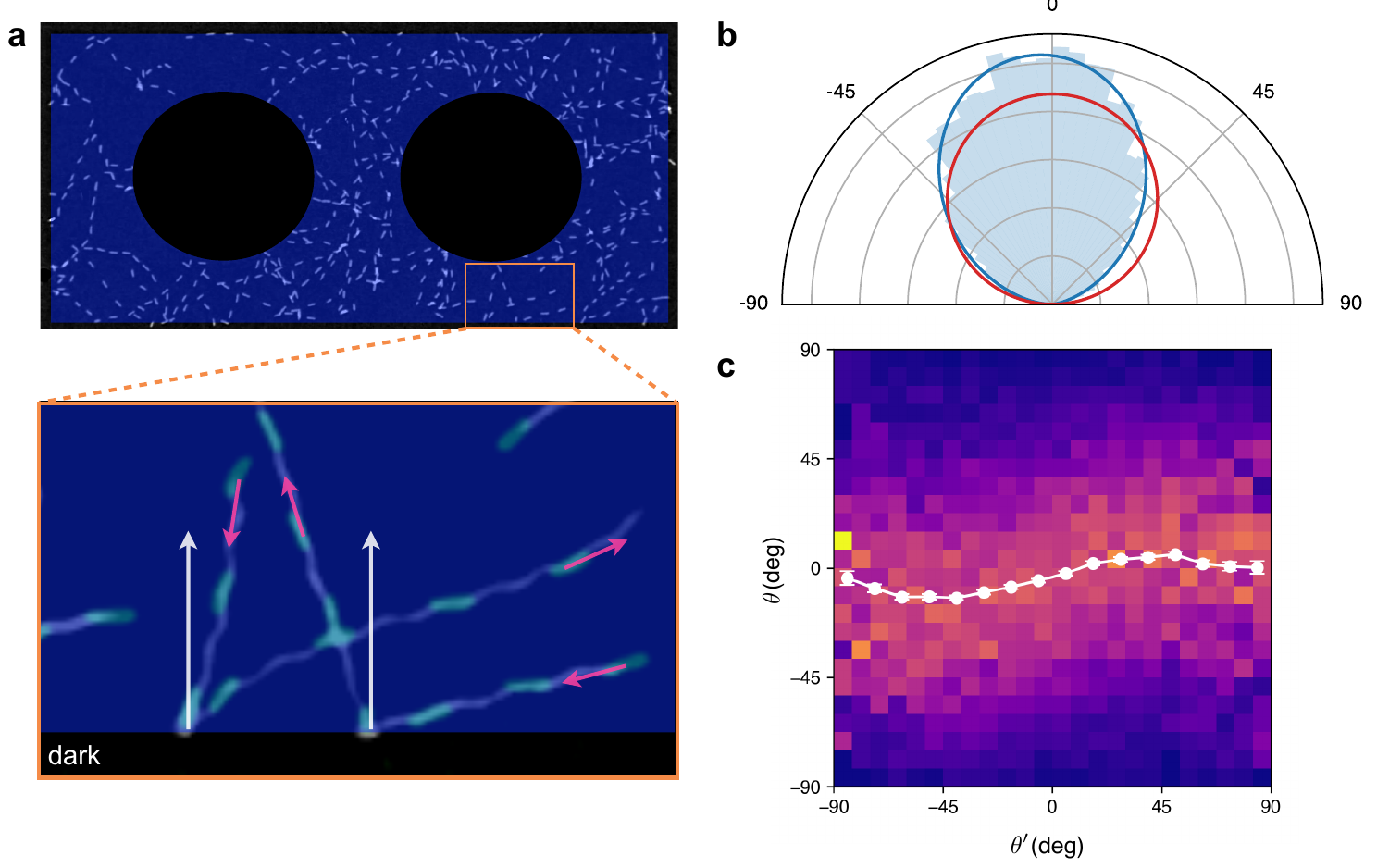}
    \caption{Scatter angle distribution of \textit{Euglena gracilis} cells. \textbf{a} Image of the measurement's light pattern with a sample timelapse of swimming cells superimposed. The enlarged area shows a zoomed view on two cells going through scattering at the light-dark boundary. White arrows depict the local boundary normal. \textbf{b} Polar histogram of the measured scattering angles. Blue line plots the best fit with $~\cos^\alpha(\theta+\theta_0)$, while the red line shows the Lambert cosine law. \textbf{c} Dependence of the outgoing angle $\theta$ from the incoming angle $\theta^\prime$. The white plot shows the mean outgoing angle in respective intervals of the incoming angle.}
    \label{fig:scatter}  
\end{figure*}

\noindent{\bf Distributions of \textit{Euglena gracilis} cells in optical confinement.}

\begin{figure*}[h]
    \centering
    \includegraphics[width=.85\textwidth]{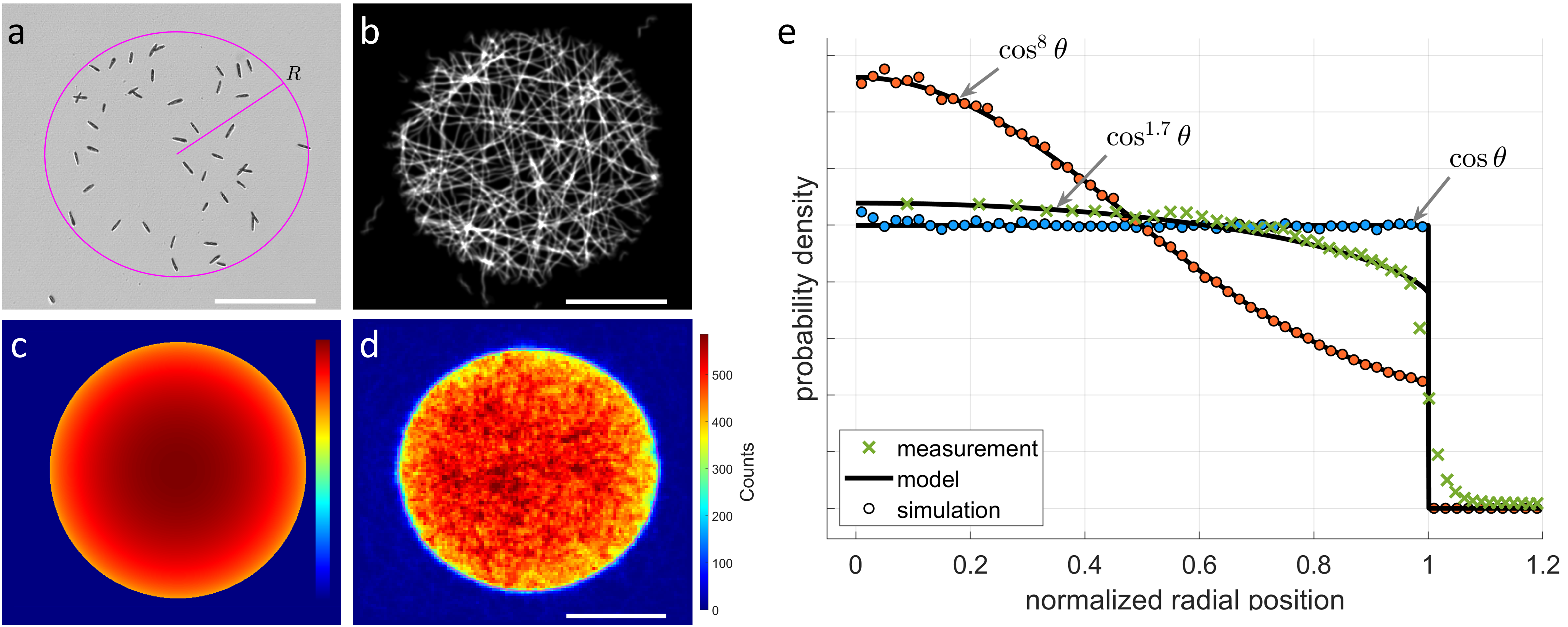}
     \caption{Spatial distribution of \textit{Euglena gracilis} cells confined in a circular light patch. \textbf{a} Bright-field image snapshot of cells during a measurement. The light pattern's edge is marked by the magenta line. \textbf{b} A random sample of persistent cell trajectories. \textbf{c} Boundary model results. \textbf{d} Two-dimensional spatial distribution of the cells (smoothed with a 0.5 bin wide Gaussian). \textbf{e} Radial spatial probability density distributions calculated from measurement data (green crosses) and from simulation results of particles with scattering distributions of $\cos \theta$ (blue circles) and $\cos^8 \theta$ (orange circles). Theoretically calculated density curves are shown as black lines. Scalebars 0.5 mm. }
    \label{fig:circle}  
   \end{figure*}

We now want to test the predictive power of our method in a real system composed of the flagellated photosynthetic alga \textit{Euglena gracilis}. \textit{Euglena} are unicellular microorganisms with an elongated body shape of approximately 50 μm in length and 10 μm diameter. As their single flagellum beats at 20-40 Hz, they swim at 50-140 μm s$^{-1}$, while simultaneously rolling at 1-2 Hz along their longitudinal axis \cite{rossi2017kinematics}. 

\textit{Euglena} cells are known to be trapped in regions of light surrounded by darkness if the illumination intensity is not too high \cite{diehn1973phototaxis}. The flagellar beat can change in response to light variations detected by a photoreceptor located at the base of the flagellum \cite{tsang2018polygonal} resulting in an inverse photophobic response at the boundary of the illuminated region, where the cells are essentially scattered back inwards. Using a blue light projection system with a properly adjusted intensity, we confined cells into two-dimensional light domains with reconfigurable geometries (see Supplementary Video 1).
Within a light pattern the cells swim in straight paths until they reach the boundary between light and dark. Here they perform a stochastic rotation until they find a new swimming direction pointing inwards. We have recorded a large number ($\sim$20000) of these scattering events within a rectangular light patch with two circular holes included to increase the probability of observing scattering events with shallow angles of incidence (Fig.\ref{fig:scatter}a)(see Supplementary Video 2). We track individual cells and analyse trajectories to fully characterize the scattering law $S(\theta^\prime,\theta)$ reported in Fig.\ref{fig:scatter}c as a density map. The outgoing angle distributions, represented by the columns in the density map, displays a weak dependence on the incoming angle $\theta^\prime$. Therefore memoryless scattering, $S(\theta^\prime,\theta)=\sigma(\theta)$, seems to be a good approximation for \textit{Euglena}. In that approximation $\sigma(\theta)$ can be directly obtained as the outgoing angle distribution averaged over all incoming angles. The result is shown as a polar histogram in Fig.\ref{fig:scatter}b together with a Lambert cosine law (red line) and a best fit representation with the model $~\cos^\alpha(\theta+\theta_0)$ (blue line, best fit parameters $\alpha=1.7$ and $\theta_0=4.2$). This scattering law is roughly intermediate between Lambert cosine and normal scattering (models 2 and 3 in Table \ref{fig:ftable}). Based on the previous discussion we would expect to find a mild accumulation in the interior of the light region. 

\begin{figure*}[h]
 \centering
 \includegraphics[width=.85\textwidth]{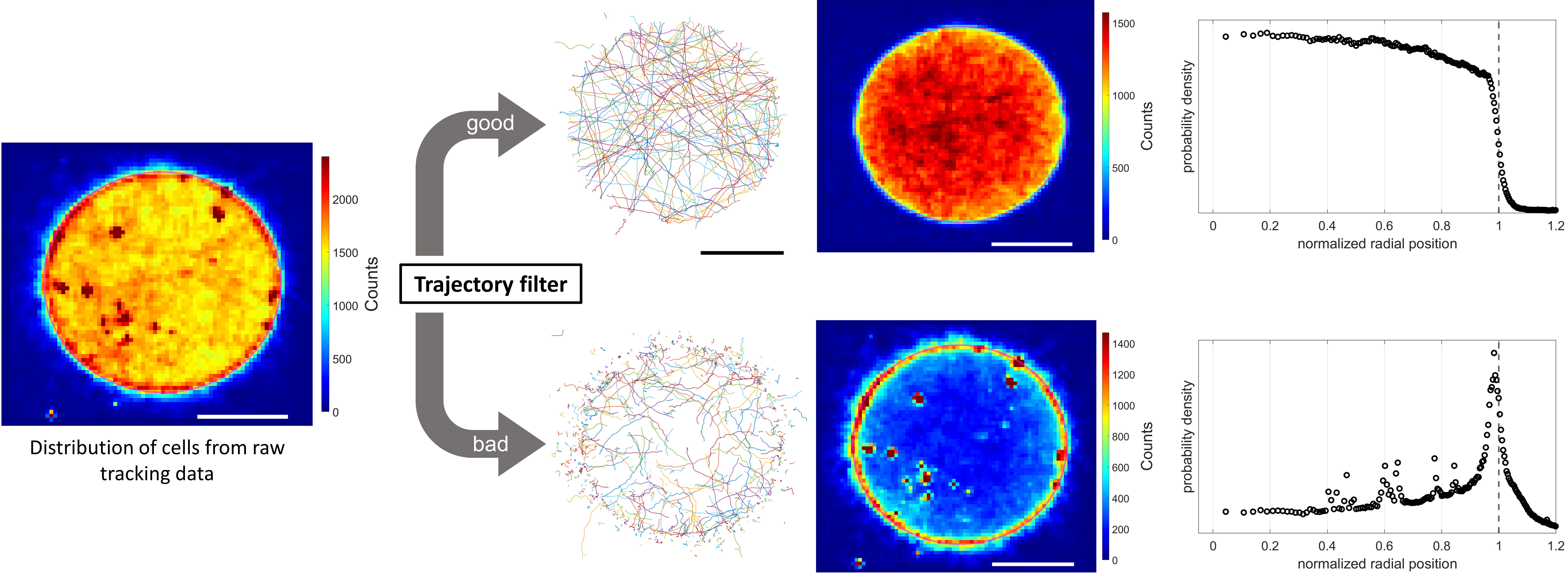}
 \caption{Demonstration of trajectory filtering. Position distribution of cells from raw tracking data is polluted by stuck cells and a background from low persistence length trajectories. A trajectory filter extracts straight trajectories from the measurements. A sample of trajectory plots, the corresponding 2D and radial cell distributions are shown for the accepted and rejected trajectory data respectively.  Scalebars 0.5 mm. }
 \label{fig:sf_circle_trajFilter} 
\end{figure*}

As a first test of our boundary method we studied \textit{Euglena} distribution inside a circular region of radius $R=630\;\mu\textrm{m}$ (Fig.\ref{fig:circle}a and Supplementary Video 3). A sample of the measured cell trajectories is shown on Fig.\ref{fig:circle}b, obtained after stuck and strongly curving trajectories were removed by a trajectory filter (see Methods and  Suppl. Fig.\ref{fig:sf_circle_trajFilter}). At a first qualitative sight, the experimental 2D density map (Fig.\ref{fig:circle}d) shows a slight depletion of cells at the boundary which compares well with the boundary method prediction shown in (Fig.\ref{fig:circle}c). For a more quantitative comparison, we make a radial histogram of cell density for which we can also derive a semi-analytical prediction from our model. Assuming isotropic dynamics, symmetry considerations imply a uniform boundary flux $p$ and a stationary density that only depends on the distance $r$ from the center. Rewriting Eq.\ref{eq:rho} in integral form we get:
\begin{equation}
    \rho(r)=
    p \int\frac{\sigma(\theta)}{d} ds=
    2 p\int_0^\pi\frac{\sigma(\theta)}{\cos\theta} d\phi
\end{equation}
where the variable $\theta$ can be expressed as a function of $\phi$ and $r$ using the condition $R\sin\theta=r\sin\phi$. For scattering laws of the form $\sigma(\theta)\propto\cos^m\theta$ we get

\begin{equation}
\rho(r)=Z\int_0^\pi\left[1-\left(\frac{r}{R}\right)^2\sin^2\phi\right]^\frac{m-1}{2} d\phi
\end{equation}
with $Z$ a normalization factor.
By numerical integration we obtain the density profiles for $n=1,8$ which agree perfectly with  simulation results as shown in Fig.\ref{fig:circle}e.  On the same figure we also show experimental data for \textit{Euglena} together with the corresponding theoretical prediction for $m=1.7$. The agreement is remarkably good in all interior points. The experimental density profile however displays a softer decay to zero at $r=R$ which we attributed to the cell's light sensing apparatus. As in many unicellular algae the photoreceptor is periodically shaded by a light absorber (eyespot) as the cell rolls during forward swimming. This produces a finite resolution in sensing the location of dark to light transition which we can estimate as the distance travelled in a half rotation during which the cell is "blind". Using average values for speed (98 $\mu$m/s) and rolling frequency (0.8 Hz), we get a distance of 62 $\mu$m, or about 0.1 R, which is enough to account for the extent of the smooth transition at the circle boundary.

{\bf Amplification of cell concentration in pattern sequences with broken spatial symmetry}.
\begin{figure*}[h]
 \centering
 \includegraphics[width=.85\textwidth]{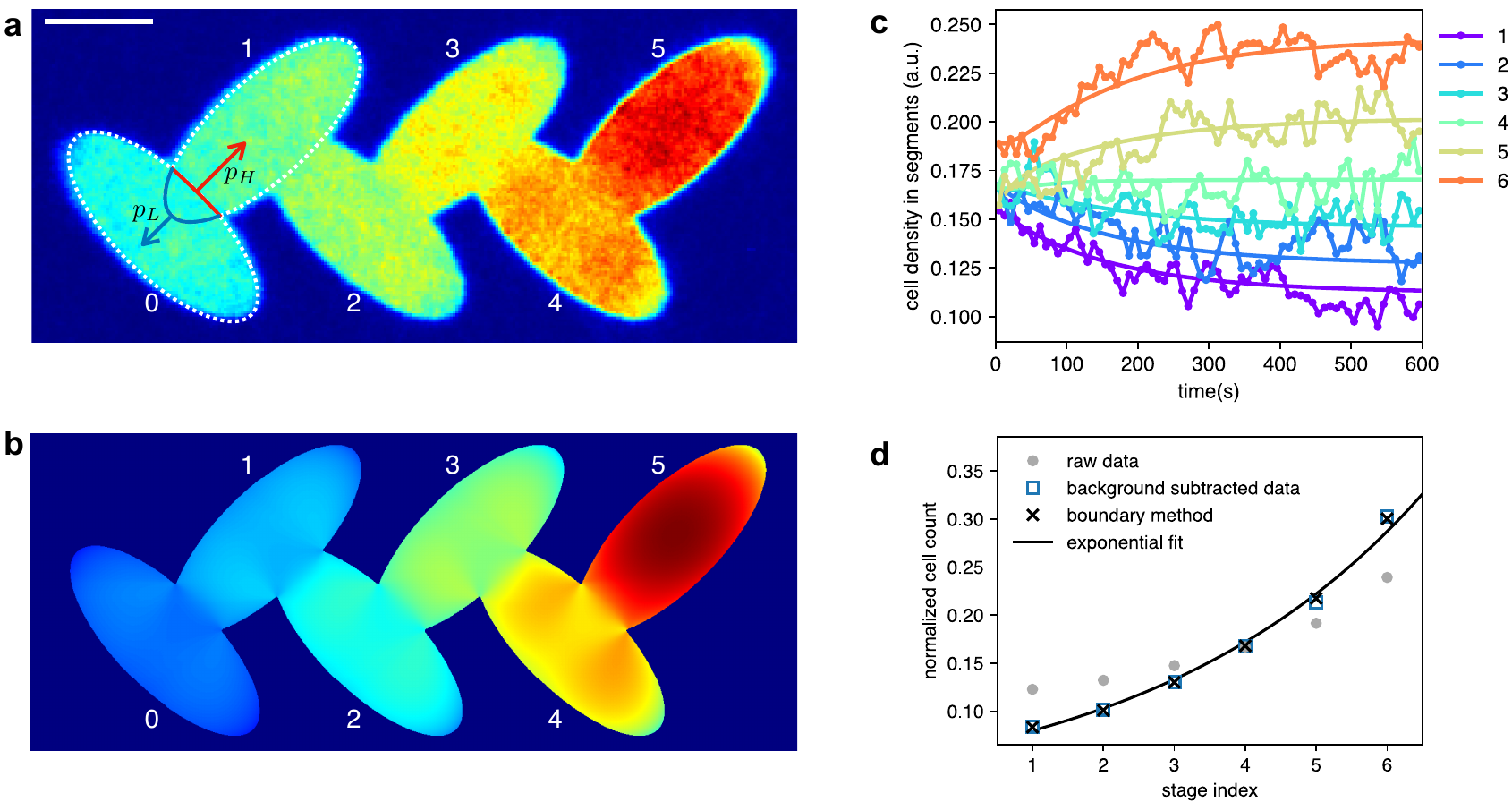}
 \caption{Rectification of \textit{Euglena gracilis} cells confined in a two-dimensional light pattern. \textbf{a} Measured steady state density map of \textit{Euglena} cells (smoothed with a 0.5 bin wide Gaussian). Scalebar 0.5 mm. \textbf{b} Boundary model result. \textbf{c} Time evolution of the normalized cell counts within the light pattern's segments (lines with markers) and their fit with Eq.\ref{eq:evo} (solid lines). \textbf{d} Normalized cell counts in the pattern's segments in steady-state (t>250s).}
 \label{fig:fratchet}  
\end{figure*}
Due to the scattering law for {\textit{Euglena}} differing only slightly from the Lambert cosine case, we observed a rather flat cell concentration profile in circular confinement. The average density within a central disk of radius $R/2$ is only 13\% higher than in the outer region. Looking at other shapes in Fig.\ref{fig:f2}, we find similar mild concentration enhancements in the central region for scattering laws close to {\textit Euglena} (row n.3 ). However, if we focus on the ellipse, we find a noticeable depletion region at the high curvature poles (left and right) compared to the low curvature poles (top and bottom). From Eq.\ref{eq:rho} we see that when we move close to a generic boundary segment $i$, local density will be dominated by the total flux $p_i$ through that $i$ segment. This consideration suggests a possible strategy to reach higher concentrations of cells by joining together equal shapes. If we start with a shape that, like the ellipse, shows an evident p variation along the boundary, we can connect a high p edge in one shape to a low p edge in a following shape, and so on. For ellipses, this means connecting them as in Fig.\ref{fig:fratchet} by repeatedly overlapping low and high curvature poles and thereby creating a sequential pattern that breaks spatial symmetry.  

We have performed experiments with two horizontally mirrored versions of the designed sequential ellipse pattern, collecting tens of millions of cell position data points. Testing the pattern in two orientations and merging the results allowed us to exclude any additional effects that might bias the cell's movement. Measurements were started with approximately uniform cell densities in each ellipse by turning on the light pattern projection over areas of homogeneously spread cells. Starting from uniformity, the cell count in each ellipse stage evolves over time until the system reaches steady state (Fig.\ref{fig:fratchet}c). Let's assume that the net flow of cells across the connections of neighboring ellipses is small enough that their spatial distribution remains close to equilibrium as the integrated cell number in each area slowly varies. Calling ${\mathbf N}(t)=\left\{{N_0,\dots,N_5}\right\}$ the vector of normalized cell counts in each area at time $t$, we can write
\begin{equation}
\dot{\mathbf N}(t)=\textsf{\textbf K}\cdot\mathbf N(t)
\end{equation}
with $\textsf{\textbf K}$ the tridiagional matrix
\begin{equation}
K_{i,i} = -(\lambda_++\lambda_-)\;,\;\;
K_{i,i+1} = \lambda_-\;,\;\;
K_{i,i-1} = \lambda_+
\end{equation}

with $\lambda_+$ and $\lambda_-$ are respectively forward and backward transition rates 
and with boundary conditions $K_{11}=-\lambda_+$, $K_{66}=-\lambda_-$.
The formal solution is
\begin{equation}
\label{eq:evo}
\mathbf N(t)=e^{\textsf{\textbf K} t}\cdot \mathbf  N(0)
\end{equation}
which can be used to fit the time series of measured normalized cell counts in Fig.\ref{fig:fratchet}c. 

In the stationary state the net flux over each junction must be zero

\begin{equation}
    N_n \lambda_+-N_{n+1}\lambda_-=0
\end{equation}
resulting in a cell count growing exponentially after each stage
\begin{equation}
    \label{eq:expo}
    N_n=N_0\left(\frac{\lambda_+}{\lambda_-}\right)^n
\end{equation}

The ratio $\lambda_+/\lambda_-$ can be estimated by our previous investigation on an isolated ellipse. Suppose we have two initially disjoint and equilibrated ellipses each one containing $N$ cells. The moment we connect them as ellipses number 0 and 1 in Fig.\ref{fig:fratchet}.a a larger number of cells will cross from 0 to 1 in unit time ($p_H$) than in the opposite direction ($p_L$). 
The values of $p_H$ and $p_L$ can be estimated as the cumulative values of $p_i\Delta_i$ over the overlapping sections of the low and high curvature boundaries of the isolated ellipses (see Fig.\ref{fig:fratchet}.a), respectively. Since $p_H\simeq N\lambda_+$ and $p_L\simeq N\lambda_-$, we obtain that $\lambda_+/\lambda_-\simeq p_H/p_L$.  


The measured steady state (t>250s) density map is shown on Fig.\ref{fig:fratchet}a. Similarly to the circle measurement, a trajectory filter was used to remove the contribution of cells that are stuck or undergoing turning at the boundary. The spatial distribution of cells is in a good match with our model's prediction (Fig.\ref{fig:fratchet}b): the cells accumulate in the predicted direction inside the confinement with matching spatial density patterns inside the individual ellipse shaped segments. For a quantitative comparison we report the normalized steady state cell count in each segment of the confinement on Fig.\ref{fig:fratchet}d. The boundary model (black crosses) seems to predict a larger concentration gradient for the cells than what found experimentally (gray circles). However if we subtract from raw data a modest (half of the smallest count) background value and renormalize (blue squares) we get an almost perfect agreement with boundary model predictions. A uniform density background can be expected due to a fraction of low persistence length cells (Suppl. Fig.\ref{fig:sf3}) and from new cells leaking into the pattern from outside, that did not had yet enough time to be effected by the confinement's shape. Renormalized data can be very well fitted by the exponential law (\ref{eq:expo}) with a fitted value for $\lambda_+/\lambda_-=1.3$. As discussed before, we can independently estimate this ratio using the boundary method results for the isolated ellipse and integrating $p$ values over the red and blue contours in Fig.\ref{fig:fratchet}a.
In this way, we obtain $p_H/p_L=1.45$, which is satisfactorily close to the measured value, considering all the approximations involved.

\section*{Discussion}

The non-equilibrium nature of active systems allows for new spatial control strategies that are not available in equilibrium systems. Here, we present a new boundary method that combines the geometric shape of the container with the particle-specific scattering law at the container walls to make accurate predictions of how active particles distribute in a confined environment.


In the specific case of \textit{Euglena} trapped within illuminated arenas, we could assume infinite persistence length and a memory-less scattering law and use our model to obtain numerical predictions very efficiently. We use our method to systematically analyze the interplay of scattering law and geometry in determining the location of concentration and depletion zones within the confined arena. Based on these results, we designed a static periodic structure specifically tailored to transport \textit{Euglena} cells through successive stages. As a result, we obtained a 3-fold concentration ratio in a six-stage structure, in excellent agreement with the predictions of the boundary method.


Our results expand our understanding of the propagation and distribution of active particles in confined spaces, allowing us to control their arrangement through the shape of the containment vessel. The presented boundary method can also serve in engineering scattering laws for micro- and macroscopic robots aiming to explore unknown spaces\cite{hidalgo2024frugal}. Extending to generic scattering laws with an explicit dependence on incident angle is possible, although calculations can be longer. A finite persistence length will play the role of volume scattering in radiometry and require information on bulk density. This could be addressed by an iterative procedure to refine the density estimate over subsequent steps. Future directions also include the possibility of designing shapes that could maximally sort active particles with different scattering laws at the boundary and shapes that generate designed flow patterns of active particles. 




\section*{Methods}

\vspace{5pt}
\noindent{\large Numerical implementation of the boundary method}\vspace{2pt}

\noindent We have implemented our boundary method in MATLAB and CUDA. We have written two CUDA GPU kernels for the calculation of the matrix $G$ and for the calculation of the in-confinement spatial probability density distributions using Eq.\ref{eq:rho}. The CUDA kernels were directly called from MATLAB using the ptx programming model. To use the CUDA GPUs best performance we have chosen to use single precision in the GPU code. A standard desktop computer was used with an NVIDIA TITAN XP GPU.

\vspace{5pt}
\noindent{\large Active particle simulation}\vspace{2pt}

\noindent In the simulation a point like particle is moved in finite steps inside a confining polygon, such that when the particle reaches the boundary, it will be scattered into a new swimming direction defined by an angle relative to the boundary normal, that is chosen randomly from a given angular probability distribution. The final results of the simulation is the spatial position distribution of the particle calculated from the accumulated position data.

\vspace{5pt}
\noindent{\large \textit{Euglena gracilis} cultures}\vspace{2pt}

Cultures were grown mixotrophically in Tris-acetate-phosphate medium \cite{gorman1965cytochrome} (TAP) in 25-mL Erlenmeyer flasks on a rotatory shaker at 130 rpm, at 23\degree C and 80 $\mu$mol photon \, m$^{-2}$ s$^{-1}$. The cultures were transferred to fresh TAP medium every two weeks, and one-week-old cultures were used for the experiments.

\vspace{5pt}
\noindent{\large \textit{Euglena gracilis} experiments}\vspace{2pt}

We have performed experiments with \textit{Euglena gracilis} in a custom built, but simple optical setup. We used a Texas Instruments DLP® LightCrafter™ DM365 digital light projector with a blue LED (470 nm) light source to project binary intensity patterns onto the sample of the cells over an 8-by-4 mm area. Illumination intensity was 0.4 \textmu W/$\mathrm{mm^2}$ to achieve positive phototactic response from the cells. Cells were placed in samples consisting of two microscope coverglasses separated by 100$\mu$m thick double sided tape. Imaging was performed under red light illumination with a Point Grey Grasshopper3 USB3 (GS3-U3-23S6M-C) camera at a magnification of 1x.

\begin{figure*}[h]
 \centering
 \includegraphics[width=.85\textwidth]{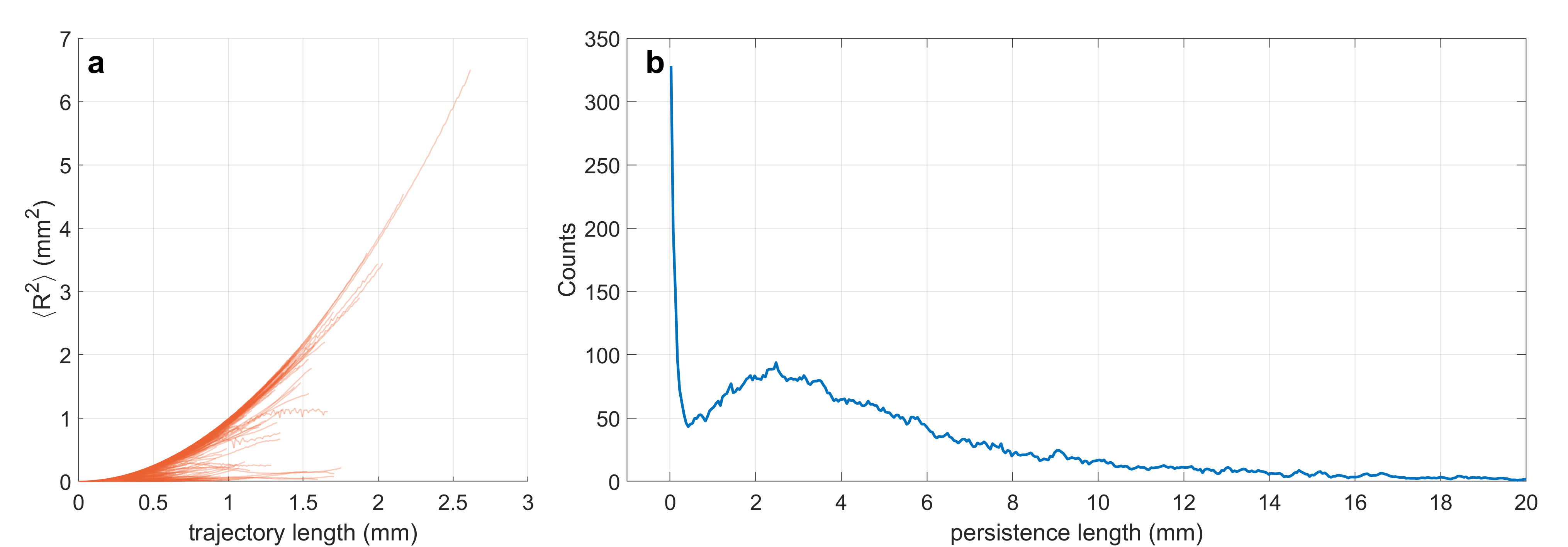}
 \caption{Persistence lengths of swimming \textit{Euglena gracilis} cells. \textbf{a} Plots of the mean squared end-to-end distance in respect to the total length for each measured trajectory. \textbf{b} Distribution of the fitted persistence lengths. About 8\% of the cells show a very low persistence length, shown by the peak in the histogram in the [0, 0.5] mm interval.}
 \label{fig:sf3} 
\end{figure*}

The persistence length of the cells was measured by tracking cells trapped in a rectangular light pattern of 4.2 x 1.5 mm size. The persistence length of the measured trajectories was obtained by fitting a Worm-like chain model:

\begin{equation}
\label{eq10}
    \left \langle R^{2} \right \rangle = 2 P L \left [ 1 - \frac{P}{L} \left ( 1 - e^{-L/P} \right ) \right ]
\end{equation}
where $\left \langle R^{2} \right \rangle$ and $L$ are the mean squared end-to-end distance and the length of a trajectory, and $P$ is the persistence length. Trajectories with lengths shorter than 0.5 mm where not considered in the analysis. Results are shown on Suppl. Fig.\ref{fig:sf3}.

\vspace{5pt}
\noindent{\large Cell tracking and trajectory filtering}\vspace{2pt}

Cells were tracked with a custom software written in MATLAB, that is capable to track cells even when they swim over each other. The recorded bright-field images were background subtracted and inverted to obtain dark background images with bright cells. After a threshold operation image objects are matched to already existing trajectories. Unmatched objects are assigned to new trajectories, unless the object's morphological parameters (area, solidity) do not match that of a single cell. Image objects that got matched to multiple trajectories are further processed as overlapping cell objects. These overlapping cells are segmented with a contour curvature segmentation algorithm designed to detect the cell's ends. From the detected cell-ends new cell positions are calculated and are than matched to the pre-matched trajectories considering both the position and the orientation of the detections and the trajectories.

To remove stuck and low persistence cells from our trajectory data we applied a trajectory filter. Here a trajectory is first segmented by its heading angle using MATLAB's ischange function. This way we are able to segment a trajectory into moving and turning segments. Then, for each segment we calculate the slope of the trajectory's direction angle, which we use to remove turning cells undergoing scattering and cells whose trajectory direction changes rapidly. To remove stuck cells we also apply a speed filter, where the mean speed of with which a trajectory segment is going away from its starting point is calculated.

For the circle measurement we used a trajectory angle slope threshold of 15 degrees/second, while for the ratchet measurements we used a value of 60 degrees/second. The speed-filter threshold was 40 $\mu$m/s for both types of measurements.

\section*{Acknowledgments}
GV was supported by the János Bolyai Research Scholarship of the Hungarian Academy of Sciences (BO/00290/21/11) and by the ÚNKP-23-5-SZTE-717 New National Excellence Program of the Ministry for Innovation and Technology from the source of The National Research, Development and Innovation Fund. This work was supported by the National Research, Development and Innovation Office, Hungary, under grant number FK 138520 and by the Hungarian Research Network under grant number SA-75/2021. RDL acknowledges funding from the European Research Council under the ERC Grant Agreement No. 834615.

\section*{Contributions}

P.O. conceived the project. A.B. and G.V. coded and run active-particle simulations. G.V and L.K. built the experimental setup. A.B., P.O. and G.V. performed experiments. D.T. and Sz.Z.T. were responsible for the growth of \textit{Euglena} cell cultures. G.V. and R.D.L analysed the experimental results. The boundary element method was created by R.D.L building upon a model developed by A.B.. G.V. wrote the boundary method's computer code. All authors contributed writing the paper.

\bibliographystyle{unsrt}
\bibliography{refs}

\begin{thebibliography}{10}

\bibitem{galajda2007wall}
Peter Galajda, Juan Keymer, Paul Chaikin, and Robert Austin.
\newblock A wall of funnels concentrates swimming bacteria.
\newblock {\em Journal of bacteriology}, 189(23):8704--8707, 2007.

\bibitem{wan2008rectification}
MB~Wan, CJ~Olson Reichhardt, Z~Nussinov, and C~Reichhardt.
\newblock Rectification of swimming bacteria and self-driven particle systems
  by arrays of asymmetric barriers.
\newblock {\em Physical review letters}, 101(1):018102, 2008.

\bibitem{lambert2010collective}
Guillaume Lambert, David Liao, and Robert~H Austin.
\newblock Collective escape of chemotactic swimmers through microscopic
  ratchets.
\newblock {\em Physical review letters}, 104(16):168102, 2010.

\bibitem{katuri2018directed}
Jaideep Katuri, David Caballero, Raphael Voituriez, Josep Samitier, and Samuel
  Sanchez.
\newblock Directed flow of micromotors through alignment interactions with
  micropatterned ratchets.
\newblock {\em ACS nano}, 12(7):7282--7291, 2018.

\bibitem{solon2015pressure}
Alexandre~P Solon, Yaouen Fily, Aparna Baskaran, Mickael~E Cates, Yariv Kafri,
  Mehran Kardar, and Julien Tailleur.
\newblock Pressure is not a state function for generic active fluids.
\newblock {\em Nature physics}, 11(8):673--678, 2015.

\bibitem{pellicciotta2023colloidal}
Nicola Pellicciotta, Matteo Paoluzzi, Dario Buonomo, Giacomo Frangipane, Luca
  Angelani, and Roberto Di~Leonardo.
\newblock Colloidal transport by light induced gradients of active pressure.
\newblock {\em Nature Communications}, 14(1):4191, 2023.

\bibitem{di2010bacterial}
Roberto Di~Leonardo, Luca Angelani, Dario Dell’Arciprete, Giancarlo Ruocco,
  Valerio Iebba, Serena Schippa, Maria~Pia Conte, Francesco Mecarini, Francesco
  De~Angelis, and Enzo Di~Fabrizio.
\newblock Bacterial ratchet motors.
\newblock {\em Proceedings of the National Academy of Sciences},
  107(21):9541--9545, 2010.

\bibitem{reichhardt2017ratchet}
CJ~Olson Reichhardt and C~Reichhardt.
\newblock Ratchet effects in active matter systems.
\newblock {\em Annual Review of Condensed Matter Physics}, 8:51--75, 2017.

\bibitem{bianchi2017holographic}
Silvio Bianchi, Filippo Saglimbeni, and Roberto Di~Leonardo.
\newblock Holographic imaging reveals the mechanism of wall entrapment in
  swimming bacteria.
\newblock {\em Physical Review X}, 7(1):011010, 2017.

\bibitem{simmchen2016topographical}
Juliane Simmchen, Jaideep Katuri, William~E Uspal, Mihail~N Popescu, Mykola
  Tasinkevych, and Samuel S{\'a}nchez.
\newblock Topographical pathways guide chemical microswimmers.
\newblock {\em Nature communications}, 7(1):10598, 2016.

\bibitem{kantsler2013ciliary}
Vasily Kantsler, J{\"o}rn Dunkel, Marco Polin, and Raymond~E Goldstein.
\newblock Ciliary contact interactions dominate surface scattering of swimming
  eukaryotes.
\newblock {\em Proceedings of the National Academy of Sciences},
  110(4):1187--1192, 2013.

\bibitem{lushi_scattering_2017}
Enkeleida Lushi, Vasily Kantsler, and Raymond~E. Goldstein.
\newblock Scattering of biflagellate microswimmers from surfaces.
\newblock {\em Physical Review E}, 96(2):023102, August 2017.

\bibitem{thery2021rebound}
Albane Th{\'e}ry, Yuxuan Wang, Mariia Dvoriashyna, Christophe Eloy, Florence
  Elias, and Eric Lauga.
\newblock Rebound and scattering of motile chlamydomonas algae in confined
  chambers.
\newblock {\em Soft Matter}, 17(18):4857--4873, 2021.

\bibitem{di2017deployable}
Raffaele Di~Giacomo, Sebastian Kr{\"o}del, Bruno Maresca, Patrizia Benzoni,
  Roberto Rusconi, Roman Stocker, and Chiara Daraio.
\newblock Deployable micro-traps to sequester motile bacteria.
\newblock {\em Scientific reports}, 7(1):45897, 2017.

\bibitem{hulme2008using}
S~Elizabeth Hulme, Willow~R DiLuzio, Sergey~S Shevkoplyas, Linda Turner,
  Michael Mayer, Howard~C Berg, and George~M Whitesides.
\newblock Using ratchets and sorters to fractionate motile cells of escherichia
  coli by length.
\newblock {\em Lab on a Chip}, 8(11):1888--1895, 2008.

\bibitem{mok2019geometric}
Rachel Mok, J{\"o}rn Dunkel, and Vasily Kantsler.
\newblock Geometric control of bacterial surface accumulation.
\newblock {\em Physical Review E}, 99(5):052607, 2019.

\bibitem{palacios2021guided}
Lucas~S Palacios, Serguei Tchoumakov, Maria Guix, Ignacio Pagonabarraga, Samuel
  S{\'a}nchez, and Adolfo G.~Grushin.
\newblock Guided accumulation of active particles by topological design of a
  second-order skin effect.
\newblock {\em Nature Communications}, 12(1):4691, 2021.

\bibitem{lam2017device}
Amy~T Lam, Karina~G Samuel-Gama, Jonathan Griffin, Matthew Loeun, Lukas~C
  Gerber, Zahid Hossain, Nate~J Cira, Seung~Ah Lee, and Ingmar~H Riedel-Kruse.
\newblock Device and programming abstractions for spatiotemporal control of
  active micro-particle swarms.
\newblock {\em Lab on a Chip}, 17(8):1442--1451, 2017.

\bibitem{steckelmacher1986knudsen}
W~Steckelmacher.
\newblock Knudsen flow 75 years on: the current state of the art for flow of
  rarefied gases in tubes and systems.
\newblock {\em Reports on Progress in Physics}, 49(10):1083, 1986.

\bibitem{ostapenko2018curvature}
Tanya Ostapenko, Fabian~Jan Schwarzendahl, Thomas~J B{\"o}ddeker,
  Christian~Titus Kreis, Jan Cammann, Marco~G Mazza, and Oliver B{\"a}umchen.
\newblock Curvature-guided motility of microalgae in geometric confinement.
\newblock {\em Physical Review Letters}, 120(6):068002, 2018.

\bibitem{cammann2021emergent}
Jan Cammann, Fabian~Jan Schwarzendahl, Tanya Ostapenko, Danylo Lavrentovich,
  Oliver B{\"a}umchen, and Marco~G Mazza.
\newblock Emergent probability fluxes in confined microbial navigation.
\newblock {\em Proceedings of the National Academy of Sciences},
  118(39):e2024752118, 2021.

\bibitem{bechinger2016active}
Clemens Bechinger, Roberto Di~Leonardo, Hartmut L{\"o}wen, Charles Reichhardt,
  Giorgio Volpe, and Giovanni Volpe.
\newblock Active particles in complex and crowded environments.
\newblock {\em Reviews of modern physics}, 88(4):045006, 2016.

\bibitem{rossi2017kinematics}
Massimiliano Rossi, Giancarlo Cicconofri, Alfred Beran, Giovanni Noselli, and
  Antonio DeSimone.
\newblock Kinematics of flagellar swimming in euglena gracilis: Helical
  trajectories and flagellar shapes.
\newblock {\em Proceedings of the National Academy of Sciences},
  114(50):13085--13090, 2017.

\bibitem{diehn1973phototaxis}
Bodo Diehn.
\newblock Phototaxis and sensory transduction in euglena: Some motile unicells
  are guided toward favorable external conditions by true homeostatic systems.
\newblock {\em Science}, 181(4104):1009--1015, 1973.

\bibitem{tsang2018polygonal}
Alan~CH Tsang, Amy~T Lam, and Ingmar~H Riedel-Kruse.
\newblock Polygonal motion and adaptable phototaxis via flagellar beat
  switching in the microswimmer euglena gracilis.
\newblock {\em Nature Physics}, 14(12):1216--1222, 2018.

\bibitem{hidalgo2024frugal}
Samuel Hidalgo-Caballero, Alvaro Cassinelli, Emmanuel Fort, and Matthieu
  Labousse.
\newblock Frugal random exploration strategy for shape recognition using
  statistical geometry.
\newblock {\em Physical Review Research}, 6(2):023103, 2024.

\bibitem{gorman1965cytochrome}
Donald~S Gorman and RP~Levine.
\newblock Cytochrome f and plastocyanin: their sequence in the photosynthetic
  electron transport chain of chlamydomonas reinhardi.
\newblock {\em Proceedings of the National Academy of Sciences},
  54(6):1665--1669, 1965.

\end{thebibliography}

\beginsupplement

\end{document}